\documentclass[useAMS,usenatbib]{mn2e}
\usepackage{graphicx}
\renewcommand{\vec}[1]{\mbox{\boldmath $#1$}}

\title[Differential rotation of main-sequence dwarfs]
{Differential rotation of main-sequence dwarfs and its
dynamo-efficiency}
\author[L.\,L.\,Kitchatinov and S.\,V.\,Olemskoy]
{L.\,L.\,Kitchatinov$^{1,2}$\thanks{E-mail: kit@iszf.irk.ru}
and S.\,V.\,Olemskoy$^{1}$ \\
$^{1}$Institute for Solar-Terrestrial Physics, PO Box 291, Irkutsk,
664033, Russia\\
$^{2}$Pulkovo Astronomical Observatory, St. Petersburg, 196140,
Russia
 }

\begin{document}

\date{Accepted .... Received ...; in original form ...}

\pagerange{\pageref{firstpage}--\pageref{lastpage}} \pubyear{2010}

\maketitle

\label{firstpage}

\begin{abstract}
A new version of a numerical model of stellar differential rotation
based on mean-field hydrodynamics is presented and tested by
computing the differential rotation of the Sun. The model is then
applied to four individual stars including two moderate and two fast
rotators to reproduce their observed differential rotation quite
closely. A series of models for rapidly rotating ($P_\mathrm{rot} =
1$~day) stars of different masses and compositions is generated. The
effective temperature is found convenient to parameterize the
differential rotation: variations with metallicity, that are quite
pronounced when the differential rotation is considered as a
function of the stellar mass, almost disappear in the dependence of
differential rotation on temperature. The differential rotation
increases steadily with surface temperature to exceed the largest
differential rotation observed to date for the hottest F-stars we
considered. This strong differential rotation is, however, found not
to be efficient for dynamos when the efficiency is estimated with
the standard $C_\Omega$-parameter of dynamo models. On the contrary,
the small differential rotation of M-stars is the most
dynamo-efficient. The meridional flow near the bottom of the
convection zone is not small compared to the flow at the top in all
our computations. The flow is distributed over the entire convection
zone in slow rotators but retreats to the convection zone boundaries
with increasing rotation rate, to consist of two near-boundary jets
in rapid rotators. The implications of the change of the flow
structure for stellar dynamos are briefly discussed.
\end{abstract}
\section{Introduction}
The theory of stellar differential rotation is mainly focused on the Sun
where helioseismology provides all the possibilities for a detailed
testing of computations. Applications to other stars are, however,
tempting in view of the rapid development of asteroseismology
\citep{C-D08} that can eventually provide data on internal stellar
rotation \citep{Sea10}. A knowledge of differential rotation is also
considered as a key for stellar dynamos.

Until recently, precise measurements of differential rotation were
mainly provided by Doppler imaging \citep{DC97,S04,Bea05}. This
technique is most suitable for rapidly rotating stars \citep{D96}.
Young dwarfs with rotation periods $P_\mathrm{rot}\sim 1$ day
present, however, difficulties for theory. Rapid rotators have
thin boundary layers at the top and bottom of their convection zones
\citep{D89,KR99}, which are difficult to resolve numerically.
Almost no computations of differential rotation for rapid rotators
were attempted.

The situation has changed recently. First, measurements of the
differential rotation of two not too rapidly rotating,
$P_\mathrm{rot}\sim 10$ days, main-sequence dwarfs were performed
using high-precision photometry of the asteroseismological {\it
MOST} mission \citep{Cea06,Wea07}. Second, a new mean-field code was
developed that can resolve boundary layers in rapidly rotating
stars; this code is used for the first time to produce the results
of this paper. We first apply it to the Sun to check its ability to
reproduce helioseismological inversions. Then, the differential
rotation of the {\it MOST}-stars ($\epsilon$~Eri and
$\kappa^1$~Ceti) are computed and compared with observations. We
also compute the differential rotation of AB Dor and LQ Hya which
probably have been the most frequent observational targets for
Doppler imaging among dwarf stars \citep{DC97,CD02,DCP03,Kea04}.

We further compute the dependence of differential rotation on the
surface temperature for rapidly rotating stars. On an observational
basis, the dependence has been studied by \citet{Bea05}. The
theoretical and observational results are quite similar. They both
show a rapid increase of surface differential rotation with
temperature. \citet{RS02,RS03} found that the rotation of F-stars
can be strongly non-uniform. Recently \citet{JD08} observed a large
differential rotation with a pole-equator lap time slightly above 10
days on a rapidly rotating G0 dwarf. Our computations suggest that
shallow convection zones of F-stars can possess even stronger
differential rotation. This raises the question of whether the
strong rotational shear implies over-normal dynamo activity. Our
computations suggest a negative answer. The efficiency of
differential rotation in generating magnetic fields can be estimated
by the modified magnetic Reynolds number that in dynamo theory is
conventionally notated as $C_\Omega$,
\begin{equation}
    C_\Omega = \frac{\Delta\Omega H^2}{\eta_{_\mathrm{T}}}
    \label{1}
\end{equation}
\citep{KR80}, where $\Delta\Omega$ is the angular velocity variation
within the convection zone, $H$ is the convection zone thickness and
$\eta_{_\mathrm{T}}$ is the turbulent magnetic diffusivity. The
$C_\Omega$ parameter (\ref{1}) is the ratio of the rate
$\Delta\Omega$ of magnetic field production by differential rotation
to the rate $\eta_{_\mathrm{T}}H^{-2}$ of diffusive escape of the
field from the convection zone. Our computations show that $C_\Omega$
{\em decreases} with stellar mass. Contrary to intuitive
expectations, a small differential rotation of M-stars is more
efficient in producing magnetic fields than the large rotational shear
of F-stars.

Differential rotation models compute the total velocity field including
the meridional flow. We discuss also the meridional flow structure
and its variations with the rotation rate.

The rest of the paper is organized as follows. Section 2 describes
our model that is based on the mean-field hydrodynamics (the
mathematical formulation is partly shifted to the Appendix). Section
3 presents and discusses the results and Section 4 summarizes the
main findings.
\section{The model}
The model is based on mean-field hydrodynamics \citep{R89}. It
computes jointly the differential rotation, meridional flow, and heat
transport in the convection zone of a star. The differential rotation of
the model results from the angular momentum transport by convection
(the $\Lambda$-effect) and meridional flow. The model is close to
its former version \citep{KR99} and will be described here only
briefly.

In order to compute differential rotation, the model needs a
knowledge of the structure of a (non-rotating) star to specify the
basic input parameters such as stellar radius, $R$, luminosity, $L$,
and mass, $M$. The structure model also supplies the density,
$\rho_\mathrm{e}$, and temperature, $T_\mathrm{e}$, at some small
depth inside the star, which depth defines the external spherical
boundary (of radius $r_\mathrm{e}$) of the simulation domain.
Displacing the external boundary by $2 - 5$ per cent in stellar
radius below the photosphere helps to avoid problems with very sharp
near-surface stratification. The reference stratification of the
model is adiabatic and spherically symmetric, and deviations from
the reference atmosphere are computed in the model. The deviations
are assumed small, so that the convection zone stratification should
be only slightly supearadiabatic and the rotation of the star should
not be too rapid, $\Omega^2R^3(GM)^{-1} \ll 1$. Using the values of
$\rho_\mathrm{e}$ and $T_\mathrm{e}$ as boundary conditions, the
equations for adiabatic profiles are integrated numerically inwards
up to the point where the radiative heat flux,
\begin{equation}
    {\vec F}^\mathrm{rad} = -\frac{16\sigma T^3}{3\kappa\rho}{\vec\nabla}T ,
    \label{2}
\end{equation}
corresponds to the total luminosity, $F^\mathrm{rad} = L(4\pi
r^2)^{-1}$. This point is the base of the convection zone. The inner
boundary ($r_\mathrm{i}$) is placed very slightly (normally by 0.1
per cent of the radius) above the base. Opacity $\kappa$ in
(\ref{2}) is computed using the {\it OPAL} opacity tables.
Therefore, the input parameters have to include the mass fraction of
Hydrogen, $X$, and metallicity, $Z$.

So defined,  the reference atmosphere helps to specify the depth profile
of the convective turnover time,
\begin{equation}
    \tau = \left( \frac{4c_\mathrm{p}\rho\ell^2 T}{3g\delta
    F}\right)^{1/3} ,
    \label{3}
\end{equation}
where $c_\mathrm{p}$ is the specific heat at constant pressure, $g$ is
gravity, $\ell = \alpha_{_\mathrm{MLT}}P(\rho g)^{-1}$ is the mixing
length, and $\delta F = L(4\pi r^2)^{-1} - F^\mathrm{rad}$ is the
\lq residual' heat flux that convection has to transport. The
turnover time (\ref{3}) enters into the key parameter of the Coriolis
number,
\begin{equation}
    \Omega^* = 2\tau\Omega .
    \label{4}
\end{equation}
The differential rotation results from the interaction between
convection and rotation \citep{L41,TT04}. The Coriolis number
(\ref{4}) measures the intensity of the interaction. The value of
$\Omega^*$ defines whether convective eddies live long enough
for the Coriolis force to affect them considerably.

The angular momentum and heat transport in the convection zone depend on the
Coriolis number. In particular, the eddy conductivity tensor,
\begin{equation}
    \chi_{ij} = \chi_{_\mathrm{T}}\left(\phi\left(\Omega^*\right)
    \delta_{ij} +  c_\chi\phi_\|\left(\Omega^*\right)
    \hat\Omega_i\hat\Omega_j \right),\ \ \
    \label{5}
\end{equation}
that controls the convective heat flux,
\begin{equation}
    F_i^\mathrm{conv} = -\rho T\chi_{ij} \frac{\partial S}{\partial r_j},
    \label{6}
\end{equation}
includes the rotationally induced anisotropy and quenching via the
functions $\phi (\Omega^*)$ and $\phi_\|(\Omega^*)$; explicit
expressions for the quenching functions are given in Kitchatinov,
Pipin \& R\"udiger (1994). In (\ref{5}) and (\ref{6}), $S$ is the
specific entropy and $\hat{\vec\Omega} = {\vec\Omega}\Omega^{-1}$ is
the unit vector along the rotation axis. Anisotropy of the eddy
conductivity (\ref{5}) means that the eddy heat flux is inclined to
the radial direction even if the entropy varies mainly in radius. As
a result, the polar regions are warmer than the equator. This \lq
differential temperature' is very important for differential
rotation models \citep{Rea05,MBT06,BR09}. It results in deviations
of the isorotational surfaces from a cylindrical shape. The
differential temperature on the Sun has been recently observed by
\citet{ROM08}.

We want to note that the eddy conductivity $\chi_{_\mathrm{T}}$ is
not prescribed, but expressed in terms of the entropy gradient,
\begin{equation}
    \chi_{_\mathrm{T}} = -\frac{\tau\ell^2 g}{12 c_\mathrm{p}}
    \frac{\partial S}{\partial r},
    \label{7}
\end{equation}
and the same for eddy viscosity. This involves an additional nonlinearity
in the equations but avoids arbitrary prescriptions of the
diffusivity profiles. The only free parameter of our model are
$c_\chi$ of (\ref{5}). The former version of the model did not use
even this parameter, instead assuming that $c_\chi = 1$ \citep{KR99}. In this
paper, we put $c_\chi = 1.5$ for closer agreement with
helioseismology.

The model solves the steady equation for the mean velocity, $\vec
u$,
\begin{equation}
    \left({\vec u}\cdot{\vec\nabla}\right){\vec u}
    + \frac{1}{\rho}{\vec\nabla}P - {\vec g} =
    -\frac{1}{\rho}{\vec\nabla}\cdot\left(\rho\hat{Q}\right),
    \label{8}
\end{equation}
together with the entropy equation,
\begin{equation}
    {\vec\nabla}\cdot\left({\vec F}^\mathrm{conv} + {\vec
    F}^\mathrm{rad}\right) + \rho T{\vec u}\cdot{\vec\nabla}S = 0.
    \label{9}
\end{equation}
In (\ref{8}), $\hat Q$ is the correlation tensor of
the fluctuating velocities ${\vec u}'$,
\begin{equation}
    Q_{ij} = \langle u'_i({\vec r}, t) u'_j({\vec r}, t)\rangle .
    \label{10}
\end{equation}
The correlation tensor can be split into a non-diffusive part,
$\hat{Q}^\Lambda$, representing the $\Lambda$-effect of non-viscous
transport of angular momentum by rotating turbulence \citep{R89}, and
the contribution $\hat{Q}^\nu$ of eddy viscosities,
\begin{equation}
    Q_{ij} = Q^\Lambda_{ij} + Q^\nu_{ij},\ \ \ Q^\nu_{ij} =-{\cal
    N}_{ijkl}\frac{\partial u_k}{\partial r_l},
    \label{11}
\end{equation}
where $\hat{\cal N}$ is the eddy viscosity tensor. The mean flow is
assumed to be axisymmetric about the rotation axis,
\begin{equation}
    {\vec u} = {\vec e}_\phi r\sin\theta\ \Omega +
    \frac{1}{\rho}{\vec\nabla}\times\left(\frac{{\vec
    e}_\phi\psi}{r\sin\theta}\right),
    \label{12}
\end{equation}
where $r,\theta,\phi$ are the usual spherical coordinates, ${\vec
e}_\phi$ is the unit vector in the azimuthal direction, and $\psi$ is
the stream function of the meridional flow. A complete representation
for the $\Lambda$-effect, eddy viscosities, and two components of
the motion equation (\ref{8}) that provide the equations for angular
velocity and meridional flow, is given in the Appendix.

The thermal condition on the top boundary assumes black-body radiation
of the photosphere. Application of the condition on the external
boundary is not straightforward, however, because of the thin
near-surface layer excluded from the simulation domain. We assume
this layer to be a perfect heat exchanger (infinite
$\chi_{_\mathrm{T}}$) so that the entropy disturbances at its base
and surface are equal. Assuming further that the disturbances are
produced at constant pressure, we find the boundary condition
\begin{equation}
    F_r = \frac{L}{4\pi r_\mathrm{e}^2}\left(1 +
    \frac{S}{c_\mathrm{p}}\right)^4\ \ \ \mathrm{at}\ \ r =
    r_\mathrm{e} .
    \label{13}
\end{equation}
$F_r$ is radial component of the total (convective plus radiative)
heat flux. The thermal condition at the inner boundary includes the
gravitational darkening effect \citep{RK02},
\begin{eqnarray}
    F_r &=& \frac{L}{4\pi r_\mathrm{i}^2}\left( 1 +
    \frac{\epsilon}{\epsilon + 3}\left( 3\cos^2\theta -
    1\right)\right) \ \ \ \mathrm{at}\ \ r = r_\mathrm{i},
    \label{14}
    \\
    \epsilon &=& \frac{\overline{\Omega}^2_\mathrm{i}
    r_\mathrm{i}}{g_\mathrm{i}},\ \ \
    \overline{\Omega}_\mathrm{i} = \frac{3}{4}\int\limits_0^\pi\
    \Omega(r_\mathrm{i},\theta )\sin^3\theta\ \mathrm{d}\theta ,
    \nonumber
\end{eqnarray}
where $\overline{\Omega}_\mathrm{i}$ is the mean angular velocity
and $g_\mathrm{i}$ is the gravity at the inner boundary.

At both boundaries, stress-free and impenetrable conditions are
applied,
\begin{equation}
    \psi = 0,\ \ Q_{r\phi} = Q_{r\theta} = 0\ \ \ \mathrm{at}\ \ r =
    r_\mathrm{i}\ \mathrm{and}\ r = r_\mathrm{e} .
    \label{15}
\end{equation}
The stress-free condition is a source of certain difficulties. The
conditions are incompatible with the Taylor--Proudman balance of the
bulk of the convection zone (the \lq thermal wind balance', in the
geophysical literature). As a result, thin layers where the balance
is violated are formed near the boundaries \citep{D89}. The boundary
layers were found in both 3D \citep{MBT06} and mean-field
simulations \citep{KR99} of differential rotation. The layer
thickness is estimated by the Eckman depth, $D\sim
\sqrt{\nu_{_\mathrm{T}}/(2\Omega )}$. In the case of rapidly
rotating stars, the layers can be very thin and difficult to resolve
numerically (the thickness $D$ decreases with angular velocity
faster than $\Omega^{-1/2}$ due to the rotational quenching of the
eddy viscosity).

To resolve the boundary layers, we apply a non-uniform grid in
radius (zeros of Chebyshev polynomials) with small spacing near the
boundaries
\begin{eqnarray}
    r_j = \frac{1}{2}\left( r_\mathrm{e} + r_\mathrm{i} -
    \left( r_\mathrm{e}-r_\mathrm{i}\right)
    \cos\left(\pi\frac{j-3/2}{n-2}\right)\right)&,&
    \\
    2 \leq j \leq n-1,\ \ \ \ \ \ \
    r_1 = r_\mathrm{i},\ \ r_n = r_\mathrm{e} ,
    \nonumber
    \label{16}
\end{eqnarray}
$n$ is the total number of grid points. For the latitude
dependencies, Legendre polynomial expansions were applied. This
leads to the two point boundary value problem in the radius for a
system of ordinary differential equations. The problem was solved
numerically by the standard relaxation method \citep{Pea92}.
\section{Results and discussion}
\subsection{Test case: the Sun}
\begin{figure}
  \resizebox{\hsize}{!}{\includegraphics{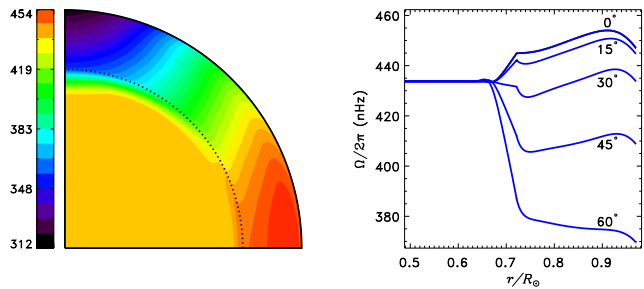}}
  \caption{Angular velocity isolines (left) and depth profiles
    of the rotation rate for several latitudes (right)
    for the model of the solar differential rotation.}
  \label{f1}
\end{figure}

Fig.~\ref{f1} shows the internal solar rotation computed with our
model. The figure includes the tachocline region and the deeper
radiation zone just for completeness of the picture. The tachocline
was computed with a separate model \citep{RK07} that uses the
results of the computation of the differential rotation of the convection zone
as a boundary condition but does not influence that computation in any
way. The results of Fig.~\ref{f1} are similar to the
helioseismological rotation law \citep{WBL97,Sea98}.

\begin{figure}
  \resizebox{\hsize}{!}{\includegraphics{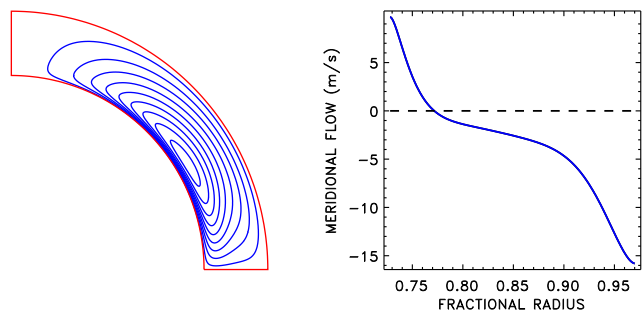}}
  \caption{Stream--lines of meridional flow (left) and radial profile
  of meridional velocity for 45{\degr} latitude (right) for the solar
  model. Negative velocity means poleward flow.}
  \label{f2}
\end{figure}

The computed meridional flow is shown in Fig.~\ref{f2}. The
direction and amplitude of the surface flow are close to
observations (Komm, Howard \& Harvey 1993). Note that the flow at
the bottom is not small compared to the surface. This is a quite
general result also found in computations for other stars. The
stagnation point in Fig.~\ref{f2} is close to the bottom so that the
downward increase of density does not lead to a slow deep
circulation. Below the convection zone, the meridional flow rapidly
decreases with depth. The distance of the flow penetration into the
tachocline is small compared to the tachocline thickness
\citep{GM04,KR06}.

The meridional flow is closely related to the Taylor--Proudman
balance. This can be seen from the meridional flow equation
\begin{equation}
    {\cal D}(\psi )\ =\ \sin\theta\ r{\partial\Omega^2\over\partial z}\
    -\ {g\over c_{\rm p} r}{\partial S\over\partial\theta},
    \label{mfeq}
\end{equation}
\citep{KR99}. In this equation, the left side describes the viscous
drag due to the meridional flow (its relation to the eddy viscosity
tensor is given by (\ref{a7}) and (\ref{11})), $\partial /\partial z
= \cos\theta\partial /\partial r - r^{-1}\sin\theta\partial
/\partial\theta$ is the spatial derivative along the rotation axis.
The two terms in the right side of (\ref{mfeq}) represent the
meridional flow driving by nonconservative parts of centrifugal and
buoyancy forces, i.e., the centrifugal and baroclinic drivings of
meridional flow, respectively. The characteristic value of each of
these two terms is large compared to the left side. Accordingly, the
two terms nearly balance each other in the bulk of convection zone.
Consequences of the Taylor--Proudman balance for the solar rotation
law have been analyzed by \citet{BRD89,BRD99}. Deviations of
isorotational surfaces from a cylindrical shape are possible due to
the latitudinal inhomogeneity of entropy, which in turn can result
from the anisotropy of the convective heat transport. The
Taylor--Proudman balance is currently rediscussed in the context of
a new hypothesis on the coincidence of the isorotational and
isentropic surfaces in rotating stars \citep{B09,Bea09}.

\begin{figure}
  \resizebox{\hsize}{!}{\includegraphics{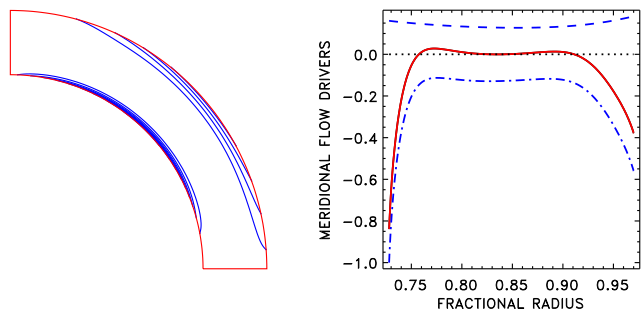}}
  \caption{
  {\sl Right:} Dependencies of the baroclinic (dashed) and
  centrifugal (dashed-dotted) terms of (\ref{mfeq}) and their
  sum (full line) on the radius for latitude 45{\degr}. The plotted
  values are normalized so that the maximum absolute value equals
  one. The dotted line shows level zero.
  {\sl Left:} Isolines of the left side of (\ref{mfeq}).
  Taylor--Proudman balance is fulfilled in the
  bulk of the convection zone but not close to the boundaries.
  }
  \label{sunbalance}
\end{figure}

Fig.~\ref{sunbalance} shows the contributions of the centrifugal and
baroclinic terms in (\ref{mfeq}), and their sum for the present
model of the solar rotation. The bulk of the convection zone is very
close to Taylor--Proudman balance. There are, however, boundary
layers where the balance is violated and the layers are not very
thin (cf.\ Balbus et al.\ 2009). Similar results on the
Taylor--Proudman balance are provided by 3D numerical simulations
(Miesch et al.\ 2006; Brun, Antia, and Chitre 2010).

In (\ref{mfeq}), we see that deviations from balance produce a
meridional flow. Observations of the global meridional flow on the
Sun  \citep{KHH93,ZK04} indicate that certain deviations from
Taylor--Proudman balance are present.
\subsection{{\it MOST}--stars}\label{MOST}
The main problem with applying the differential rotation model to
individual stars is to specify the (input) stellar parameters. Of the
two stars -- $\epsilon$ Eridani and $\kappa^1$ Ceti -- whose
differential rotation was measured using the {\it MOST}-data
\citep{Cea06,Wea07}, $\epsilon$ Eri presents much less difficulties
because all the required parameters were estimated by \citet{SD89}.

We used the EZ code of stellar evolution by \citet{P04} to define
the structure of a main-sequence star of given mass, age, and
composition and infer the input parameters for our simulations from
the structure model. Hydrogen content was fixed to $X=0.7$. The
parameters used to model the differential rotation of {\it
MOST}-stars are given in Table~\ref{t1}. The parameters of
$\epsilon$~Eri given by \citet{SD89} can be closely reproduced by
the structure model of $0.8M_\odot$ star with metallicity $Z=0.01$
and an age of 1~Gyr. The parameters of $\kappa^1$~Ceti are less
certain. Those used in differential rotation measurements
\citep{Rea04,Wea07} can be roughly reproduced by the structure model
of a $1M_\odot$ star with $Z=0.02$ at the age of about 600~Myr.

\begin{table}
    \caption{Input parameters of the differential rotation models
    for {\it MOST}-stars.}
    \label{t1}
    \begin{tabular}{@{}lcccccc}
    \hline
    Star & $M/M_\odot$ & $R/R_\odot$ & $L/L_\odot$ & $Z$ & Age, Gyr &
    $P_\mathrm{rot}$ \\
    \hline
    $\epsilon$ Eri  & 0.8 & 0.724 & 0.337 & 0.01 & 1 & 11 \\
    $\kappa^1$ Ceti & 1.0 & 0.907 & 0.758 & 0.02 & 0.6 & 9 \\
    \hline
    \end{tabular}

    \medskip
    $P_\mathrm{rot}$ is in days.
\end{table}

\begin{figure}
  \resizebox{\hsize}{!}{\includegraphics{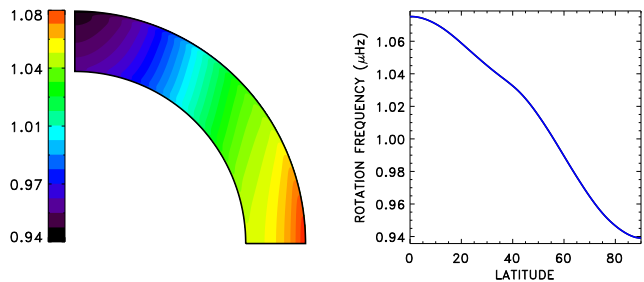}}
  \caption{Angular velocity isolines (left) and surface profile of
    the rotation rate (right) for the differential rotation model
    of $\epsilon$~Eridani.}
  \label{f3}
\end{figure}
\begin{figure}
  \resizebox{\hsize}{!}{\includegraphics{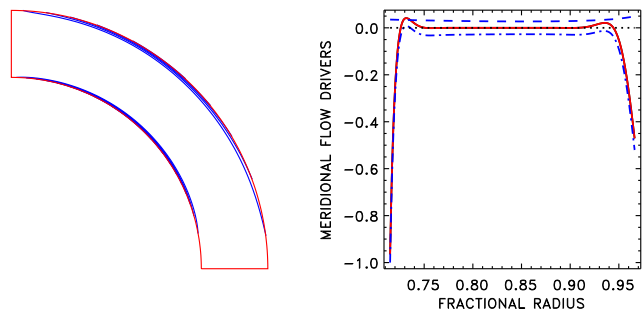}}
  \caption{
  Same as in Fig.~\ref{sunbalance} but for the differential rotation
  model of $\epsilon$~Eridani.
  }
  \label{eribalance}
\end{figure}
\begin{figure}
  \resizebox{\hsize}{!}{\includegraphics{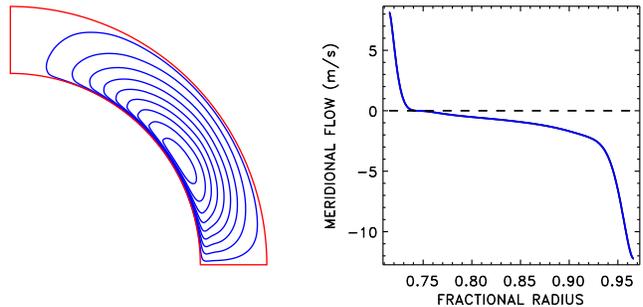}}
  \caption{Simulated meridional flow of $\epsilon$~Eridani. Stream
  lines are shown in the left panel and the right panel shows the depth
  profile of the meridional velocity for latitude 45{\degr}.}
  \label{f4}
\end{figure}

Fig.~\ref{f3} shows the results of differential rotation
simulation for $\epsilon$~Eri. The relative magnitude of the surface
differential rotation can be estimated with the parameter
\begin{equation}
    \alpha_{_\mathrm{DR}} = 1 -
    \frac{\Omega_\mathrm{pole}}{\Omega_\mathrm{eq}} .
    \label{17}
\end{equation}
Our model gives the value of $\alpha_{_\mathrm{DR}} = 0.127$ for
$\epsilon$~Eri, close to the observational value of
$\alpha_{_\mathrm{DR}} = 0.11$ \citep{Cea06}. The agreement for
$\kappa^1$~Ceti is not so close: $\alpha_{_\mathrm{DR}} = 0.130$ is the
computational value and $\alpha_{_\mathrm{DR}} = 0.09$ is the result of measurements
\citep{Wea07}. The simulated rotation laws for both \lq moderate
rotators' are quite similar. The dependence of the rotation rate on
the latitude in Fig.~\ref{f3} is not as smooth as for the solar model.
There is a \lq peculiarity' in the surface profile located around
the latitude where the angular velocity isoline tangential to the
inner boundary at the equator arrives at the surface. We always find
such a peculiarity in rotation laws computed for stars rotating
considerably faster than the Sun. This means that the often used
approximation
\begin{equation}
    \Omega = \Omega_\mathrm{eq}\left( 1 -
    \alpha_{_\mathrm{DR}}\cos^2\theta\right)
    \label{18}
\end{equation}
may not be very accurate. This peculiarity also means that moderate
rotators are much closer to the strict Taylor--Proudman balance than
the Sun. This balance is illustrated by Fig.~\ref{eribalance}. The
centrifugal and baroclinic terms in the meridional flow equation
(\ref{mfeq}) strictly balance each other everywhere except for the
thin boundary layers. Violation of this balance in the layers
excites a meridional flow. Accordingly, the meridional flow of
Fig.~\ref{f4} is highly concentrated at the boundaries. The bottom
flow is not small compared to the top but the flow in the bulk of
the convection zone away from the boundary layers is slow. We shall
see that the boundary layers are even more pronounced in rapid
rotators.

There is an increasing belief that meridional flow is important for
solar \citep{CSD05} and stellar \citep{JBB10} dynamos. The flow
structure prescribed in advection-dominated dynamo models is,
however, different from the flow predicted by stellar
circulation models.
\subsection{Rapid rotators}
Simulations of differential rotation were performed for two young
stars -- AB~Doradus and LQ~Hydrae -- that seem to be the most frequent
observational targets among the rapid rotators. Other dwarf stars
for which the differential rotation was measured by Doppler imaging
are either not yet settled on the main sequence or their structure
parameters are difficult to determine. The parameters used in
differential rotation simulations that also help to reproduce
closely the observational structure parameters of AB~Dor
\citep{DC97,Oea07,GMM08} and LQ~Hya \citep{Kea04} are listed in
Table~\ref{t2}.

\begin{table}
    \caption{Input parameters of the differential rotation models
    for AB Doradus and LQ Hydrae.}
    \label{t2}
    \begin{tabular}{@{}lcccccc}
    \hline
    Star & $M/M_\odot$ & $R/R_\odot$ & $L/L_\odot$ & $Z$ & Age &
    $P_\mathrm{rot}$ \\
    \hline
    AB Dor  & 0.9 & 0.803 & 0.438 & 0.02 & 70 & 0.514 \\
    LQ Hya & 0.77 & 0.698 & 0.273 & 0.01 & 100 & 1.6 \\
    \hline
    \end{tabular}

    \medskip
    Age is given in Myr, $P_\mathrm{rot}$ -- in days.
\end{table}

\begin{figure}
  \resizebox{\hsize}{!}{\includegraphics{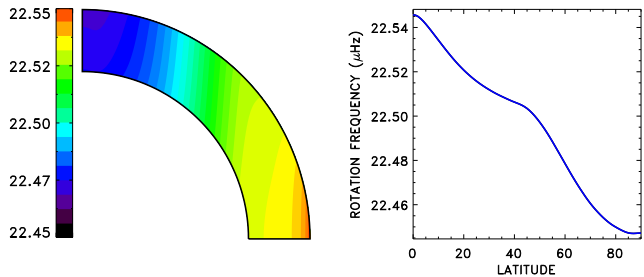}}
  \caption{Angular velocity isolines (left) and latitude dependence of
    the surface rotation frequency (right) for the differential rotation model
    of AB Dor.}
  \label{f5}
\end{figure}
\begin{figure}
  \resizebox{\hsize}{!}{\includegraphics{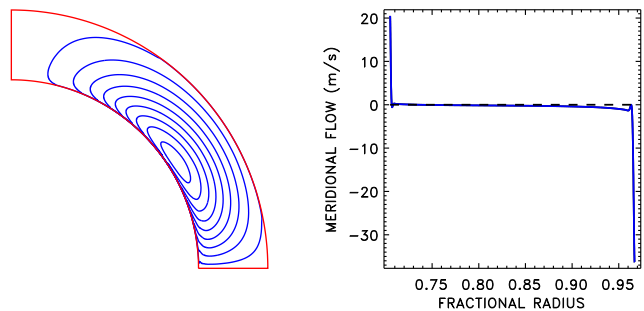}}
  \caption{Simulated meridional flow of AB Dor. The left panel shows the
  stream lines and the right panel presents the depth profile of
  meridional velocity for the latitude of 45$^\circ$. Negative velocity
  means poleward flow.}
  \label{f6}
\end{figure}

Figs.~\ref{f5} and \ref{f6} show the modelled differential rotation
and meridional flow of AB~Dor. The computed differential rotation
measure $\alpha_{_\mathrm{DR}} = 4.37\times 10^{-3}$ is very close
to the observational value of $4.5\times 10^{-3}$ \citep{DC97}. The
peculiarity in the surface profile of rotation rate discussed in
Section~\ref{MOST} is even more pronounced in Fig.~\ref{f5} compared
to the moderate rotation case of Fig.~\ref{f3}. The profile can be
only roughly approximated by the $\cos^2\theta$-law of
(\ref{18}). If the approximation is nevertheless used to
describe the rotation of the stellar spots, it may lead to a seeming variation
of the differential rotation with time. The spots positioned at
different latitudes for different observational epochs would lead to
different $\alpha_{_\mathrm{DR}}$ suggesting torsional oscillations
even for a steady rotation law.

Observational estimates of the differential rotation of the slower
rotating LQ Hya have a wide spread \citep{Bea05}. We can only state
that the differential rotation measure of our model,
$\alpha_{_\mathrm{DR}} = 1.28\times 10^{-2}$, is within the range of
observational estimates.

The meridional flow of Fig.~\ref{f6} shows an extreme concentration
in the boundary layers. The flow consists of two near-boundary jets
linked by a very slow circulation in the bulk of the convection
zone. Such a boundary-layer flow is, probably, not important for
dynamos. However, the distributed (solar-type) flow of Fig.~\ref{f2}
may be significant for magnetic field transport. The meridional
circulation changes from a distributed flow (Fig.~\ref{f2}) to the
near-boundary jets (Fig.~\ref{f6}) with an increasing rotation rate.
This change of meridional flow may cause a change in the dynamo
regime that may be the reason for the two separate branches for fast
and slow rotators in the dependence of the dynamo-cycle period on
the stellar rotation rate found by \citet{SB99}.
\subsection{Temperature dependence}
\begin{figure}
  \resizebox{\hsize}{!}{\includegraphics{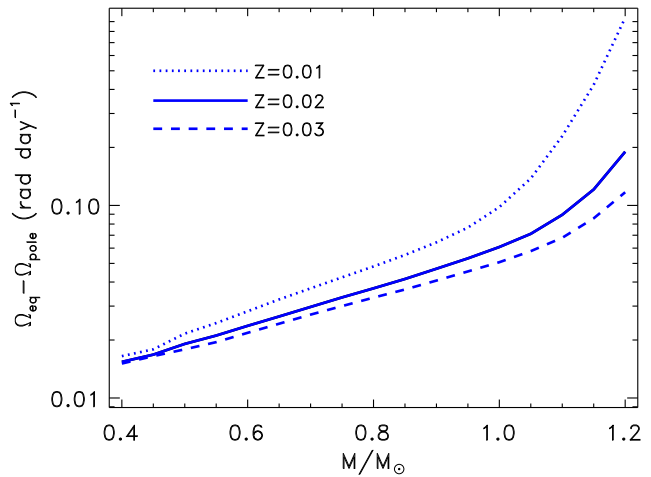}}
  \caption{Surface differential rotation of {\it ZAMS}-stars
    with $P_\mathrm{rot} = 1$~day as a function of stellar mass.
    Three lines show the results of computations for metallicities
    $Z = 0.01$ (dotted), $Z=0.02$ (full line), and $Z=0.03$ (dashed).}
  \label{f7}
\end{figure}

Fig.~\ref{f7} shows the dependence of the surface differential
rotation on stellar mass computed with our model. The computations
were made for young stars just arrived on the main sequence and
rotating with a period of 1~day. Models were produced for the mass
range from $0.4M_\odot$ to $1.2M_\odot$ (with $0.05M_\odot$
spacing). These computations cover the surface temperature range from
about 3600~K to 6500~K or spectral types from K2 to F6 which roughly
corresponds to the range for which \citet{Bea05} constructed the
temperature dependence of the surface differential rotation detected
by Doppler imaging. The computations were made for three
metallicity values of $Z = 0.01,\ 0.02,\ 0.03$. For a given stellar
mass, the results depend on the chemical composition, so that for a
mass of $1.2M_\odot$, the surface differential rotation differs by a
factor of about 10 between the cases of $Z=0.01$ and $Z=0.03$.

\begin{figure}
  \resizebox{\hsize}{!}{\includegraphics{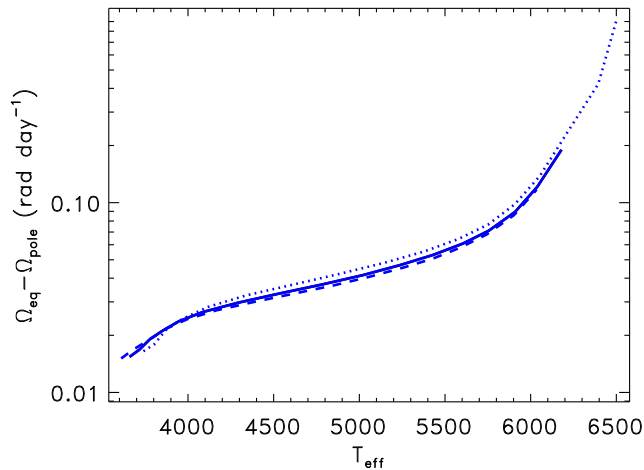}}
  \caption{Same differential rotation as in Fig.~\ref{f7} but shown
    as a function of surface temperature.
    The results for different metallicities of
    $Z = 0.01$ (dotted line), $Z=0.02$ (full line), and $Z=0.03$
    (dashed line) now differ by little.}
  \label{f8}
\end{figure}

The metallicity dependence almost disappear, however, when the
differential rotation is plotted as a function of surface
temperature (Fig.~\ref{f8}). So, the effective temperature is indeed
convenient for parameterizing the differential rotation of young stars
\citep{Bea05}. An even better scaling can be found by using the
Coriolis number of (\ref{4}). The results for different chemical
compositions practically coincide on the plot of differential
rotation as a function of the Coriolis number. As the Coriolis number
is not directly observable, we shall keep using the surface
temperature.

\begin{figure}
  \resizebox{\hsize}{!}{\includegraphics{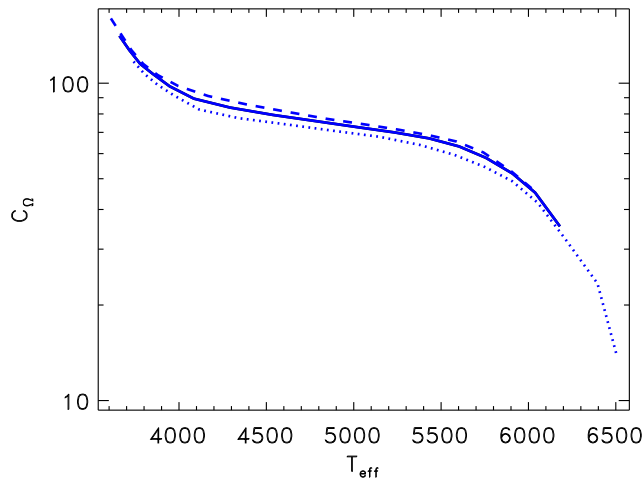}}
  \caption{$C_\Omega$ dynamo-number of (\ref{1}) as a function of surface
      temperature. The lines of different styles show the results for
      different metallicities, $Z=0.01$ (dotted), $Z=0.02$ (full
      line), and $Z=0.03$ (dashed).}
  \label{f9}
\end{figure}

Figs. \ref{f7} and \ref{f8} predict that relatively hot
convective stars can possess strong differential rotations with
pole-equator lap times shorter than 10 days. This is larger than the
strongest differential rotation observed to date \citep{JD08}. The
question arises whether a strong differential rotation implies
over-normal dynamo-activity. The results of Fig.~\ref{f9} suggest a
negative answer. The figure shows the $C_\Omega$ dynamo-number of
(\ref{1}) as a function of surface temperature.

The $C_\Omega$-parameter was computed for the middle of the convection
zone using the isotropic part of the eddy magnetic diffusivity
$\eta_{_\mathrm{T}} = \chi_{_\mathrm{T}}\phi(\Omega^*)$ (that
coincides with the eddy conductivity; \citealt{KPR94}) and
latitudinal differential rotation (the radial inhomogeneity of rotation
is relatively small). The estimation assumes that dynamos of young
stars are distributed over their convection zones \citep{D99}.

The $C_\Omega$ of Fig.~\ref{f9} declines sharply with temperature
for F-stars indicating that the strong differential rotation of
these stars is not efficient at producing toroidal magnetic fields.
This is in agreement with the idea of \citet{DL78} that convective
dynamos cease to operate at about spectral type F6. The $C_\Omega$
increases steadily with decreasing temperature. This is because the
convection slows down in low mass stars to decrease eddy diffusion.
The decline of magnetic eddy diffusivity overpowers the decrease of
differential rotation to produce a larger $C_\Omega$ in smaller stars.
The largest $C_\Omega$ belong to M-dwarfs. This is in contrast to
the common belief that the small differential rotation of M-stars
cannot be important for dynamos and that the magnetic fields of these stars
are generated by the $\alpha^2$-mechanism. The $\alpha^2$ dynamos
produce nonaxisymmetric global fields. However, observations favour
an axial symmetry of the global magnetic structure of M-stars
\citep{Dea06}. This may be explained by the effect of differential
rotation, which is small in low-mass stars, but efficient in winding
magnetic fields.

The gravitational darkening of (\ref{14}) is not significant for our
results. Neglecting the darkening effect reduces the computed
differential rotation normally by less than 1 per cent (by several
per cent in extreme cases of rapidly rotating F-stars).
\section{Summary}
This paper presents the first results of the new mean-field model of stellar
differential rotation, which improves on its former formulation
\citep{KR99}, to cover the case of rapidly rotating stars with
$P_\mathrm{rot}\sim 1$~day.

The model reproduces very closely helioseismological inversions for
the internal solar rotation. The simulated meridional flow at the
bottom of the solar convection zone has an amplitude of about
10~m\,s$^{-1}$ that is not small compared to the surface flow. The
near-bottom equator-ward flow can be important for the solar dynamo.

The meridional flow in stars rotating faster than the Sun is
increasingly concentrated in boundary layers near the top and bottom
of the convection zone as the rotation rate increases. We interpret
this boundary-layer structure of the meridional flow as an effect of
thin boundary layers where Taylor--Proudman balance is violated. The
change of the meridional flow from distributed to boundary-layer
structure may be the reason for the change of dynamo regime between
slow and fast rotators \citep{SB99}.

The differential rotation model was applied to four individual stars
including two moderate ($P_\mathrm{rot}\sim 10$~days) and two fast
($P_\mathrm{rot}\sim 1$~day) rotators. In two cases, for which the
structure parameters of the stars are well known, close agreement
with observations was found. In all cases, the computed rotation
laws were not so close to the $\sin^2(\mathrm{latitude})$-profile of
equation (\ref{18}) as it is for the Sun.

The computations for the rapidly rotating ($P_\mathrm{rot} = 1$~day) {\it
ZAMS}-stars show that the surface temperature, $T_\mathrm{eff}$, is
a convenient parameter for the differential rotation: when
considered as a function of $T_\mathrm{eff}$, the differential
rotation loses its dependence on the chemical composition of a star that
otherwise can be quite pronounced. The differential rotation
increases with $T_\mathrm{eff}$ and the rotation rate difference
between equator and poles can reach almost 1~rad\,day$^{-1}$ for the
hottest F-stars we considered.

This strong differential rotation is, however, not efficient for
dynamos. The standard $C_\Omega$-parameter of the dynamo models of
equation (\ref{1}) that measures the efficiency of toroidal field
production by differential rotation {\em decreases} with
$T_\mathrm{eff}$. Contrary to intuitive expectation, the small
differential rotation of M-stars is important for magnetic field
generation. This may be the reason for the closeness of the observed
magnetic structure of M-stars to axial symmetry \citep{Dea06}.

As a perspective for future work, theoretical construction of the
dependence of differential rotation on stellar age and temperature
based on girochronology \citep{B08} can be pointed out.
\section*{Acknowledgments}
This work was supported by the Russian Foundation for Basic Research
(projects 10-02-00148, 10-02-00391).

\appendix
\section{Motion equations}
\subsection{Reynolds stress}
The Reynolds stress tensor is related to the fluctuating velocity
correlation of (\ref{10}) and (\ref{11}), $R_{ij} =-\rho
Q_{ij}$. The part $Q^\Lambda_{ij}$ of the correlation tensor, which
represents the $\Lambda$-effect of the non-viscous transport of angular
momentum in stratified rotating fluids, reads
\begin{eqnarray}
    Q^\Lambda_{ij} &=&
    \nu_{_\mathrm{T}}\left(\frac{\alpha_{_\mathrm{MLT}}}{\gamma}\right)^2
    \bigg( \left(J_0(\Omega^*) + a I_0(\Omega^*)\right) \times
    \nonumber \\ &&
    \left(\hat{r}_i\varepsilon_{jkl}+\hat{r}_j\varepsilon_{ikl}\right)
    - \left(J_1(\Omega^*) + a I_1(\Omega^*)\right)\times
    \nonumber \\ &&
    \frac{\left(\hat{\vec r}\cdot{\vec\Omega}\right)}{\Omega^2}
    \left(\Omega_i\varepsilon_{jkl} +\Omega_j\varepsilon_{ikl}\right)\bigg)
    \Omega_k\hat{r}_l ,
    \label{a1}
\end{eqnarray}
where $\hat{\vec r}$ is the radial unit vector, $a$ is a parameter
of convection anisotropy ($a=2$ in all our computations), and
$\gamma = c_\mathrm{p}/c_\mathrm{v}$. Recent discussions of the
$\Lambda$-effect can be found in \citet{RK07} and \citet{Gea10}. The
origin of the expression (\ref{a1}) for the $\Lambda$-effect was
discussed by \citet{KR05} where expressions for the functions $J_0,\
J_1,\ I_0,\ I_1$ of the Coriolis number $\Omega^*$ are also given.

The viscous part of the Reynolds stress is controlled by the
viscosity tensor of (\ref{11}). The viscosity is anisotropic due
to the rotational influence on turbulent convection,
\begin{eqnarray}
    {\cal N}_{ijkl} &=& \nu_{_\mathrm{T}}\big(
    \phi_1(\Omega^*)\left( \delta_{ik}\delta_{jl}\ +\
    \delta_{jk}\delta_{il}\right) \nonumber \\
    &+& \phi_2(\Omega^*)\big( \delta_{il}\hat\Omega_j\hat\Omega_k +
    \delta_{jl}\hat\Omega_i\hat\Omega_k \nonumber \\
    &+&\delta_{ik}\hat\Omega_j\hat\Omega_l +
    \delta_{jk}\hat\Omega_i\hat\Omega_l +
    \delta_{kl}\hat\Omega_i\hat\Omega_j \big) \nonumber \\
    &+& \phi_3(\Omega^*)\ \delta_{ij}\delta_{kl}\ -\
    \phi_4(\Omega^*)\ \delta_{ij}\hat\Omega_k\hat\Omega_l
    \nonumber \\
    &+& \phi_5(\Omega^*)\ \hat\Omega_i\hat\Omega_j\hat\Omega_k\hat\Omega_l
    \big) .
    \label{a2}
\end{eqnarray}
The viscosity quenching functions, $\phi_n(\Omega^*),\ n=1,...,5$,
can be found in \citet{KPR94}. The eddy viscosity
$\nu_{_\mathrm{T}}$ for a non-rotating fluid is expressed in terms of
the entropy gradient
\begin{equation}
    \nu_{_\mathrm{T}} = -\frac{\tau\ell^2 g}{15 c_\mathrm{p}}
    \frac{\partial S}{\partial r}.
    \label{a3}
\end{equation}
\subsection{Angular velocity equation}
The azimuthal component of (\ref{8}) gives the continuity equation
for the angular momentum flux,
\begin{eqnarray}
    \frac{1}{r^2\rho}\frac{\partial}{\partial
    r}\left(\rho r^3 Q_{r\phi}\right) +
    \frac{1}{\sin^2\theta}\frac{\partial}{\partial\theta}\left(\sin^2\theta\
    Q_{\theta\phi}\right) &+&
    \nonumber \\
    \frac{1}{r^2\rho}\frac{\partial(r^2\Omega )}{\partial
    r}\frac{\partial\psi}{\partial\theta} -
    \frac{1}{\rho\sin^2\theta}\frac{\partial(\sin^2\theta\ \Omega
    )}{\partial\theta}\frac{\partial\psi}{\partial r} &=& 0 ,
    \label{a4}
\end{eqnarray}
where the first and the second lines describe angular momentum
transport by convection and meridional flow respectively. On using
(\ref{11}) and (\ref{a1}) -- (\ref{a3}), the convective fluxes
of angular momentum can be written as follows
\begin{eqnarray}
    Q_{\theta\varphi } &=&
    \sin\theta\ {\tau\ell^2 g\over 15 c_\mathrm{p}}
    {\partial S\over\partial r}\bigg\{ \phi_1(\Omega^*)
    {\partial\Omega\over\partial\theta} \nonumber \\
    &-&\phi_2(\Omega^* ) \sin\theta \left( \cos\theta\ r
    {\partial\Omega\over\partial r }\ -\ \sin\theta
    {\partial\Omega\over\partial\theta }\right)
    \nonumber\\
    &-& \Omega \left({\alpha_{\rm MLT}\over\gamma}\right)^2
    \sin\theta\cos\theta\ \left(J_1(\Omega^*) + a I_1(\Omega^*)\right)
    \bigg\} ,
    \nonumber \\
    Q_{r\varphi } &=&
    \sin\theta\ {\tau\ell^2 g\over 15 c_\mathrm{p}}
    {\partial S\over\partial r}\bigg\{ \phi_1(\Omega^*)
    r{\partial\Omega\over\partial r} \nonumber \\
    &+&\phi_2(\Omega^* ) \cos\theta \left( \cos\theta\ r
    {\partial\Omega\over\partial r }\ -\ \sin\theta
    {\partial\Omega\over\partial\theta }\right)
    \nonumber\\
    &-& \Omega \left({\alpha_{_\mathrm{MLT}}\over\gamma}\right)^2
    \big( J_0(\Omega^*) + a I_0(\Omega^*)
    \nonumber \\
    &-& \cos^2\theta\ \left(J_1(\Omega^*) + a I_1(\Omega^*)\right)\big) \bigg\} .
    \label{a5}
\end{eqnarray}
With the angular momentum fluxes (\ref{a5}), (\ref{a4})
governs the angular velocity distribution in the convection zone.
\subsection{Meridional flow equation}
The equation (\ref{mfeq}) for the meridional flow can be found as
the azimuthal component of the curled equation (\ref{8}). The left
part,
\begin{equation}
    {\cal D}(\psi ) = \varepsilon_{\phi jk}\frac{\partial}{\partial
    r_j}\left( \frac{1}{\rho}\frac{\partial\left(\rho Q^\nu_{kl}\right)}{\partial
    r_l}\right),
    \label{a7}
\end{equation}
of this equation describes the viscous drag to the meridional flow. In
spherical coordinates, (\ref{a7}) reads
\begin{eqnarray}
    {\cal D}(\psi ) &=& \frac{1}{r}\frac{\partial}{\partial r} \left(
    \frac{1}{\rho r^2}\frac{\partial (\rho r^3
    Q^\nu_{r\theta})}{\partial r}\right)
    \nonumber \\
    &-&
    \frac{1}{r^2}\frac{\partial}{\partial\theta}\left(\frac{1}{\sin\theta}\frac{\partial\left(\sin\theta\
    Q^\nu_{r\theta}\right)}{\partial\theta}\right)
    \nonumber \\
    &-& \frac{1}{r\rho}\frac{\partial\rho}{\partial r}\frac{\partial
    Q^\nu_{rr}}{\partial\theta}
    +\frac{1}{r}\frac{\partial^2}{\partial r\partial\theta}\left( Q^\nu_{\theta\theta}-Q^\nu_{rr}\right)
    \nonumber \\
    &+& \frac{\cos\theta}{r \sin\theta}\frac{\partial}{\partial
    r}\left( Q^\nu_{\theta\theta}-Q^\nu_{\phi\phi}\right)
    \nonumber \\
    &+& \frac{1}{r^2}\frac{\partial}{\partial\theta}\left(Q^\nu_{\theta\theta}
    + Q^\nu_{\phi\phi} - 2 Q^\nu_{rr}\right) .
    \label{a8}
\end{eqnarray}
The explicit expression for ${\cal D}(\psi )$ in terms of the stream
function is very complicated and never used in practice. Instead,
$Q^\nu_{r\theta}$, $u_\theta$ and a certain combination of diagonal
components of $Q^\nu_{ij}$ are introduced as new dependent variables
when solving the meridional flow equation. The expressions for these
new variables in terms of the stream function can by found from
(\ref{11}), (\ref{12}), and (\ref{a2}).

\label{lastpage}


\begin{thebibliography}{99}
\bibitem[\protect\citeauthoryear{Balbus}{2009}]{B09}
    Balbus S.A.,
    2009, MNRAS, 395, 2056
\bibitem[\protect\citeauthoryear{Balbus et al.}{2009}]{Bea09}
    Balbus S.A., Bonart J., Latter H.N., Weiss N.O.,
    2009, MNRAS, 400, 176
\bibitem[\protect\citeauthoryear{Barnes}{2008}]{B08}
    Barnes S.A.,
    2008, IAUS, 258, 345
\bibitem[\protect\citeauthoryear{Barnes et al.}{2005}]{Bea05}
    Barnes J.R., Collier Cameron A., Donati J.-F., James D.J.,
    Marsden S.C., Petit P.,
    2005, MNRAS, 357, L1
\bibitem[\protect\citeauthoryear{Brun \& Rempel}{2009}]{BR09}
    Brun A.S., Rempel M.,
    2009, SSRv, 144, 151
\bibitem[\protect\citeauthoryear{Brun, Antia \& Chitre}{2010}]{BAC10}
    Brun A.S., Antia H.M., Chitre S.M.,
    2010, A\&A, 510, 33
\bibitem[\protect\citeauthoryear{Choudhuri, Sch\"ussler \& Dikpati}{1995}]{CSD05}
    Choudhuri A.R., Sch\"ussler M., Dikpati M.,
    1995, A\&A, 303, L29
\bibitem[\protect\citeauthoryear{Christensen-Dalsgaard}{2008}]{C-D08}
    Christensen-Dalsgaard J.,
    2008, IAUS, 252, 135
\bibitem[\protect\citeauthoryear{Collier Cameron \& Donati}{2002}]{CD02}
    Collier Cameron A., Donati J.-F.,
    2002, MNRAS, 329, L23
\bibitem[\protect\citeauthoryear{Croll et al.}{2006}]{Cea06}
    Croll B., Walker G.A.H., Kuschnig R. et al.,
    2006, ApJ, 648, 607
\bibitem[\protect\citeauthoryear{Donati}{1996}]{D96}
    Donati J.-F.,
    1996, IAUS, 176, 53
\bibitem[\protect\citeauthoryear{Donati}{1999}]{D99}
    Donati J.-F.,
    1999, MNRAS, 302, 457
\bibitem[\protect\citeauthoryear{Donati \& Collier Cameron}{1997}]{DC97}
    Donati J.-F., Collier Cameron A., 1997, MNRAS, 291, 1
    1996, IAUS, 176, 53
\bibitem[\protect\citeauthoryear{Donati, Collier Cameron \& Petit}{2003}]{DCP03}
    Donati J.-F., Collier Cameron A., Petit P.,
    2003, MNRAS, 345, 1187
\bibitem[\protect\citeauthoryear{Donati et al.}{2006}]{Dea06}
    Donati J.-F., Forveille T., Collier Cameron A., Barnes J.R.,
    Delfosse X., Jardine M.M., Valenti J.A.,
    2006, Sci, 311, 633
\bibitem[\protect\citeauthoryear{Durney}{1989}]{D89}
    Durney B.R.,
    1989, ApJ, 338, 509
\bibitem[\protect\citeauthoryear{Durney}{1999}]{BRD99}
    Durney B.R.,
    1999, ApJ, 511, 945
\bibitem[\protect\citeauthoryear{Durney \& Latour}{1978}]{DL78}
    Durney B.R., Latour J.,
    1978, GApFD, 9, 241
\bibitem[\protect\citeauthoryear{Garaud et al.}{2010}]{Gea10}
    Garaud P., Ogilvie G.I., Miller N., Stellmach S.,
    2010, arXiv:astro-ph/1004.3239
\bibitem[\protect\citeauthoryear{Gilman \& Miesch}{2004}]{GM04}
    Gilman P.A., Miesch M.S.,
    2004, ApJ, 611, 568
\bibitem[\protect\citeauthoryear{Guirado Marti-Vidal \& Marcaide}{2008}]{GMM08}
    Guirado J.C., Marti-Vidal I., Marcaide J.M.,
    2008, IAUS, 248, 496
\bibitem[\protect\citeauthoryear{Jeffers \& Donati}{2008}]{JD08}
    Jeffers S.V., Donati J.-F.,
    2008, MNRAS, 390, 635
\bibitem[\protect\citeauthoryear{Jouve, Brown \& Brun}{2010}]{JBB10}
    Jouve L., Brown B.P., Brun A.S.,
    2010, A\&A, 509, 32
\bibitem[\protect\citeauthoryear{Kitchatinov \& R\"udiger}{1999}]{KR99}
    Kitchatinov L.L., R\"udiger G.,
    1999, A\&A, 344, 911
\bibitem[\protect\citeauthoryear{Kitchatinov \& R\"udiger}{2005}]{KR05}
    Kitchatinov L.L., R\"udiger G.,
    2005, Astron. Nachr., 326, 379
\bibitem[\protect\citeauthoryear{Kitchatinov \& R\"udiger}{2006}]{KR06}
    Kitchatinov L.L., R\"udiger G.,
    2006, A\&A, 453, 329
\bibitem[\protect\citeauthoryear{Kitchatinov et al.}{1994}]{KPR94}
    Kitchatinov L.L., Pipin V.V., R\"udiger G.,
    1994, Astron. Nachr., 315, 157
\bibitem[\protect\citeauthoryear{Komm et al.}{1993}]{KHH93}
    Komm R.W., Howard R.F., Harvey J.W.,
    1993, SoPh, 147, 207
\bibitem[\protect\citeauthoryear{K\H{o}v\'{a}ri et al.}{2004}]{Kea04}
    K\H{o}v\'{a}ri Z., Strassmeier K.G., Granzer T., Weber M.,
    Ol\'{a}h~K., Rice J.B.,
    2004, A\&A, 417, 1047
\bibitem[\protect\citeauthoryear{Krause \& R\"adler}{1980}]{KR80}
    Krause F., R\"adler K.-H.,
    1980, Mean-Field Magnetohydrodynamics
    and Dynamo Theory. Akademieverlag, Berlin
\bibitem[\protect\citeauthoryear{Lebedinsky}{1941}]{L41}
    Lebedinsky A.I.,
    1941, Astron. Zh., 18, 10
\bibitem[\protect\citeauthoryear{Ortega et al.}{2007}]{Oea07}
    Ortega V.G., Jilinski E., de La Reza R., Bazzanella B.,
    2007, MNRAS, 377, 441
\bibitem[\protect\citeauthoryear{Miesch, Brun \& Toomre}{2006}]{MBT06}
    Miesch M.S., Brun A.S., Toomre J.,
    2006, ApJ, 641, 618
\bibitem[\protect\citeauthoryear{Paxton}{2004}]{P04}
    Paxton B.,
    2004, PASP, 116, 699
\bibitem[\protect\citeauthoryear{Press et al.}{1992}]{Pea92}
    Press W.H., Teukolsky S.A., Vetterling W.T., Flannery B.P.,
    1992, Numerical Recipes. Cambridge Univ. Press, Cambridge
\bibitem[\protect\citeauthoryear{Rast, Ortiz \& Meisner}{2008}]{ROM08}
    Rast M.P., Ortiz A., Meisner R.W.,
    2008, ApJ, 673, 1209
\bibitem[\protect\citeauthoryear{Reiners \& Schmitt}{2002}]{RS02}
    Reiners A., Schmitt J.H.M.M.,
    2002, A\&A, 393, L77
\bibitem[\protect\citeauthoryear{Reiners \& Schmitt}{2003}]{RS03}
    Reiners A., Schmitt J.H.M.M.,
    2003, A\&A, 398, 647
\bibitem[\protect\citeauthoryear{Rucinski et al.}{2004}]{Rea04}
    Rucinski S.M., Walker G.A.H., Matthews J.M. et al.,
    2004, PASP, 116, 1093
\bibitem[\protect\citeauthoryear{R\"udiger}{1989}]{R89}
    R\"udiger G.,
    1989, Differential Rotation and Stellar Convection. Gordon \&
    Breach, New York
\bibitem[\protect\citeauthoryear{R\"udiger \& Kitchatinov}{2007a}]{RK07}
    R\"udiger G., Kitchatinov L.L.,
    2007, NJPh, 9, 302
\bibitem[\protect\citeauthoryear{R\"udiger \& Kitchatinov}{2007b}]{RK07b}
    R\"udiger G., Kitchatinov L.L.,
    2007, in Hughes D.W., Rosner R., Weiss N.O., eds, The Solar
    Tachocline. Cambridge Univ. Press, Cambridge, p.\ 129
\bibitem[\protect\citeauthoryear{R\"udiger \& K\"uker}{2002}]{RK02}
    R\"udiger G., K\"uker M.,
    2002, A\&A, 385, 308
\bibitem[\protect\citeauthoryear{R\"udiger et al.}{2005}]{Rea05}
    R\"udiger G., Egorov P., Kitchatinov L.L., K\"uker M.,
    2005, A\&A, 431, 345
\bibitem[\protect\citeauthoryear{Saar \& Brandenburg}{1999}]{SB99}
    Saar S.H., Brandenburg A.,
    1999, ApJ, 524, 259
\bibitem[\protect\citeauthoryear{Schou et al.}{1998}]{Sea98}
    Schou J., Antia H.M., Basu S. et al.,
    1998, ApJ, 505, 390
\bibitem[\protect\citeauthoryear{Soderblom \& D\"appen}{1989}]{SD89}
    Soderblom D.R., D\"appen W.,
    1989, ApJ, 342, 945
\bibitem[\protect\citeauthoryear{Strassmeier}{2004}]{S04}
    Strassmeier K.G.,
    2004, IAUS, 219, 11
\bibitem[\protect\citeauthoryear{Su\'arez et al.}{2010}]{Sea10}
    Su\'arez J.C., Andrade L., Goupil M.J., Janot-Pacheco E.,
    2010, arXiv:astro-ph/1004.0609
\bibitem[\protect\citeauthoryear{Tassoul \& Tassoul}{2004}]{TT04}
    Tassoul J.-L., Tassoul M.,
    2004, A Concise Hystory of Solar and Stellar Physics.
    Princeton Univ. Press, Princeton, NJ, p.\ 215
\bibitem[\protect\citeauthoryear{Walker et al.}{2007}]{Wea07}
    Walker G.A.H., Croll B., Kuschnig R. et al.,
    2007, ApJ, 659, 1611
\bibitem[\protect\citeauthoryear{Wilson, Burtonclay \& Li}{1997}]{WBL97}
    Wilson P.R., Burtonclay D., Li Y.,
    1997, ApJ, 489, 395
\bibitem[\protect\citeauthoryear{Zhao \& Kosovichev}{2004}]{ZK04}
    Zhao J., Kosovichev A.G.,
    2004, ApJ, 603, 776
\end{thebibliography}
\end{document}